\documentclass[iop]{emulateapj}
\usepackage{amsmath,amssymb}
\usepackage{multirow}
\usepackage{xcolor}
\usepackage{natbib}
\usepackage[colorlinks,citecolor=blue,linkcolor=blue,urlcolor=black]{hyperref}
\usepackage{etoolbox}

\makeatletter

\patchcmd{\NAT@citex}
  {\@citea\NAT@hyper@{%
     \NAT@nmfmt{\NAT@nm}%
     \hyper@natlinkbreak{\NAT@aysep\NAT@spacechar}{\@citeb\@extra@b@citeb}%
     \NAT@date}}
  {\@citea\NAT@nmfmt{\NAT@nm}%
   \NAT@aysep\NAT@spacechar\NAT@hyper@{\NAT@date}}{}{}

\patchcmd{\NAT@citex}
  {\@citea\NAT@hyper@{%
     \NAT@nmfmt{\NAT@nm}%
     \hyper@natlinkbreak{\NAT@spacechar\NAT@@open\if*#1*\else#1\NAT@spacechar\fi}%
       {\@citeb\@extra@b@citeb}%
     \NAT@date}}
  {\@citea\NAT@nmfmt{\NAT@nm}%
   \NAT@spacechar\NAT@@open\if*#1*\else#1\NAT@spacechar\fi\NAT@hyper@{\NAT@date}}
  {}{}

\makeatother

\begin{document}               % plus the \end{document} command at the end.

%\setcounter{secnumdepth}{4}
%\renewcommand{\subsubsubsection}{\paragraph} 

%\makeatletter
%\renewcommand\subsubsection{\@startsection{subsubsection}{3}{\z@}%
%                                     {-3.25ex\@plus -1ex \@minus -.2ex}%
%                                     {1.5ex plus .2ex}%
%                                     {\normalfont\normalsize}}
%\renewcommand\subsubsubsection{\@startsection{paragraph}{4}{\z@}%
%                                     {-1.25ex\@plus -1ex \@minus -.2ex}%
%                                     {0.0001pt \@plus .2ex}%margin
%                                     {\normalfont\normalsize\it}}
%\makeatother

\newcommand{\specialcell}[2][c]{%
  \begin{tabular}[#1]{@{}c@{}}#2\end{tabular}}

\mathchardef\mhyphen="2D % Define a "math hyphen"

%\title{The Local \& Universal LGRB Rate Explained}
%\title{Explaining the Local \& Universal LGRB Rate with Metallically}
\title{The Absolute Rate of LGRB Formation}

\author{J. F. Graham$^{\hyperref[jfg]{1}}$ \affil{Max-Planck Institute for Extraterrestrial Physics, 85748 Garching, Germany}}
\author{P. Schady$^{\hyperref[jfg]{1}}$ \affil{Max-Planck Institute for Extraterrestrial Physics, 85748 Garching, Germany}}
\affil{$^1$\label{jfg} Max-Planck Institute for Extraterrestrial Physics, Giessenbachstrasse 1, 85748 Garching, Germany}
%\author{A. S. Fruchter\affil{Space Telescope Science Institute, 3700 San Martin Drive, Baltimore MD 21218}}
%, Giessenbachstrasse 1

%\author{\today}
\journalinfo{}
\submitted{}

\begin{abstract}

We estimate the LGRB progenitor rate using our recent work on the effects of environmental metallically on LGRB formation in concert with SNe statistics via an approach patterned loosely off the Drake equation.  Beginning with the cosmic star-formation history, we consider the expected number of broad-line Type Ic events (the SNe type associated with LGRBs) that are in low metallicity host environments adjusted by the contribution of high metallicity host environments at a much reduced rate.  We then compare this estimate to the observed LGRB rate corrected for instrumental selection effects to provide a combined estimate of the efficiency fraction of these progenitors to produce LGRBs and the fraction of which are beamed in our direction.  From this we estimate that an aligned LGRB occurs for approximately every 4000 low metallically broad-lined Type Ic Supernovae.  Therefore if one assumes a semi-nominal beaming factor of 100 then only about one such supernova out of 40 produce an LGRB.  Finally we propose an off-axis LGRB search strategy of targeting for radio observation broad-line Type Ic events that occur in low metallicity hosts.

\end{abstract}

\section{Introduction}

This is the second of two papers where we directly look at the effect of environmental metallically on the Long-duration Gamma Ray Burst (LGRB) rate.  In the first paper \cite{diff_rate_letter} we did a relative comparison of the impact of host galaxy metallically on the rate of LGRBs per unit star-formation available.  Here we look at the absolute LGRB rate using our knowledge of the effects of metallically on LGRB formation and the established  connection between LGRBs and broad-lines Type Ic (Ic-bl) SNe \citep{Modjaz2008}.

Since the BeppoSAX detection of LGRB 980425 coincident in space and time with Type Ic supernova SN 1998bw (and to convince the doubters the detection of LGRB 030329 with an optical afterglow on the exact position of its supernova) we have known that at least some LGRBs have an associated supernova event \citep{Pian980425, Galama980425b}. However it was immediately obvious that not all supernovae (SNe) could result in LGRBs as there were nowhere near enough LGRBs for all the supernovae even after correcting for the factor of LGRBs that were not beamed in our direction.  As the population of LGRB SNe grew it became clear that they where not just associated with a particular type of SNe, Ic, but that the LGRB associated events had unusually broad spectral features \citep{Modjaz2008}.  However without considering the impact of metallicity on the LGRB formation rate there remained too many of the specific broad-lined Type Ic SNe compared with too few LGRBs, even given the crude estimate of the latter.  Even assuming that the $\sim$30\% of dark LGRBs \citep{Perley_dusty_LGRBs} are  systematically overlooked due their lack of redshift information (and therefore assumed to be at a very high redshift not used for comparison) does not change this discrepancy between Ic-bl SNe and LGRB numbers in any significant way.

If we know the rate of LGRBs relative to other core-collapse events, then we can study the mechanisms required for LGRB formation.
%Furthermore if the rate of LGRBs is known observationally the accuracy of such an understanding can be assessed.
  Three recent developments allow for a significant advance in this approach: \cite{Lien} has exhaustively analyzed the {\it Swift}-BATÕs performance to estimate the true GRB rate, \cite{stats_paper} show that observations of LGRBs mostly occurring in low metallicity environments is a strong intrinsic preference; and \cite{diff_rate_letter} quantized this as most LGRBs occurring in the least metal rich $\sim$10\% of the star-formation with LGRB events at higher metallicallies occurring less than 25 times as often per unit star-formation. % Here we will use these three pieces to probe the LGRB formation puzzle.

Now with a much improved understanding of the LGRB rate and of the impact of environmental metallicity on LGRB formation we will conduct a more detailed analysis.  Here we will begin, in section \ref{drake}, with the Cosmic Star-Formation Rate (CSFR) History, correlate this to the supernova rate, step though the selections necessary for a supernova to also give rise to an LGRB.  In section \ref{rate_comp}, we compare this predicted rate with the observed LGRB rate corrected for instrumental selection effects.  We assess the difference between our analytically derived LGRB rate and observationally-based LGRB rate, and what it tells us about LGRB progenitors.

% observational estimate the LGRB rate, and assess the remaining gap between the number of LGRBs and the number of possible progenitor events.

%accuracy of and constrains on our understanding of this process.

% Here we will begin with the SNe rate, step though the selections necessary for a supernova to also give rise to an LGRB, thus estimate the LGRB rate, compare it with observation, and assess the accuracy of and constrains on our understanding of this process.

%\section{The Road to an LGRB}
\section{A Drake Equation for LGRBs} \label{drake}

Our analysis will begin with, $R_{SF}$, the Cosmic Star-Formation Rate (CSFR).  Star-formation has the advantage of being studied both for the local universe and across the universe's history.  Since LGRB progenitors have short lifetimes, LGRBs trace the on-going star formation, and their cosmic rate should thus match the CSFR after taking into account the conditions necessary to form an LGRB.
%Since LGRB progenitors are expected to have lifetimes shorter than the timescale for any significant variability in the SFR the currently observable distribution of star-formation should match the LGRB distribution without an observable lag after accounting for the process necessary to form an LGRB.
  Estimates of the current CSFR, $R_{SF}$(z=0), are in reasonable agreement 0.0158$^{+0.003}_{-0.004}$ \citep{Hanish}, 0.0193 $\pm$ 0.0012 (\citealt{Horiuchi} analysis of \citealt{Bothwell} data), 0.01729 $\pm$ 0.0035 (average of \citealt{Bothwell} Table 3 literature values) M$_\sun$ yr$^{-1}$ Mpc$^{-3}$.
% both internally and with the z = 0 values to attempts to fit the CSFR evolution as a function of redshift.  
We thus adopt a value of 0.0175 $\pm$ 0.0025 for $R_{SF}$(z=0).

To extend the CSFR as a function of redshift we adopt the piecewise \cite{hb} model.  This model fits the CSFR with a series of three power-law estimates across different redshift ranges roughly corresponding to 0 $<$ z $<$ 1,  1 $<$ z $<$ 4.5, and z $>$ 4.5 (see Table \ref{hb_val}).  We believe the initial and middle pieces of this estimate to be reasonably well constrained observationally with the final z $\gtrsim$ 4.5 term being highly uncertain (even \citealt{hb} increases there uncertainty in this region to $>$50\%).  \cite{Yuksel} have attempted to use LGRBs to probe probe the CSFR at z $\gtrsim$ 4 and suggest a much reduced rate of decline and thus higher CSFR than estimated in \cite{hb}.  The \cite{hb} (and other conventional) CSFR ÔestimatesÕ are based on fits to deep, multi-band galaxy observations, and the difference between CSFR at z $>$ 4 derived from LGRBs and galaxy observations could be the result of a significant fraction of star-formation at z $>$ 4 occurring in low-luminosity, low-mass galaxies that are undetected in traditional galaxy surveys.
However, \cite{Robertson2012} claim that such an LGRB determined high redshift SFR would overpredict the observed stellar mass density and thus the LGRB production rate (at least at z $>$ 4) may be effected by factors other than SFR and environmental metallicity.  To avoid these issues altogether we will look only at the CSFR in the initial and middle piecewise power-law terms.  We adopt the functional CSFR form of \cite{Yuksel} along with the $\eta$ = -10 smoothing term, as shown in equation \ref{hb_eqn}, but to prevent circular reasoning (i.e. using an LGRB based CSFR to estimate the LGRB rate) we retain the \cite{hb} values as shown in Table \ref{hb_val} (this distinction is actually relatively trivial as we limit our comparison to avoid the high redshift range where the estimates diverge).

\begin{figure*}[!t]
\begin{equation}
\label{hb_eqn}
R_{SF}(z) = R_{SF}(z=0) \Bigg[ \bigg(1+z \bigg)^{a\eta} + \bigg({1+z \over{(1+z_1)^{1-a/b}}}\bigg)^{b\eta} + \bigg({1+z \over{(1+z_1)^{(b-a)/c}} (1+z_2)^{1-b/c}}\bigg)^{c\eta}  \Bigg]^{1/\eta}
\end{equation}
%\vspace*{-0.5 cm}
\end{figure*}

%\begin{equation}
%R_{SF}(z) = R_{SF}(z=0) \Bigg[ \bigg(1+z \bigg)^{a\eta} + \bigg({(1+z)^b \over{(1+z_1)^{b-a}}}\bigg)^{\eta} + \bigg({(1+z)^c \over{(1+z_1)^{b-a}} (1+z_2)^{c-b}}\bigg)^{\eta}  \Bigg]^{1/\eta}
%\end{equation}

%\begin{equation}
%R_{SF}(z) = R_{SF}(z=0) \Bigg[ (1+z)^{a\eta} + {(1+z)^{b\eta} \over{(1+z_1)^{(b-a)\eta}}} + {(1+z)^{c\eta} \over{(1+z_1)^{(b-a)\eta}} (1+z_2)^{(c-b)\eta}}  \Bigg]^{1/\eta}
%\end{equation}

\begin{table}[h!]
\vspace{-0.3 cm}
\caption{\label{hb_val} CSFR Parameters}
\vspace{-0.4 cm}
\begin{center}
\begin{tabular}{cc}
\hline
\hline
Variable & Value \\
\hline
$R_{SF}$(z=0) & 0.0178 \\
a &  3.28 \\
b & -0.26 \\
c & -8.0 \\
z$_1$ &  1.04  \\
z$_2$ &  4.48 \\
$\eta$ & -10 \\
\hline
\end{tabular}
\end{center}
\vspace{-0.2 cm}
{ \cite{hb} values for equation \ref{hb_eqn}.  Note: With the exception of $R_{SF}$(z=0) and $\eta$, these are a subset of the \cite{SalpeterIMF} IMF values in \cite{hb} Table 2 (of these only the independent parameters needed for equation \ref{hb_eqn} are shown).} 
\end{table}

Then we will consider the steps necessary to convert star-formation to LGRBs.% and estimate fraction of events, $f$, for each step. 
As it is often more convenient to think in numbers that are greater than 1 we will throughout this work refer to the reciprocals of these fractions as factors, $f^{-1}$.  For instance if a nominal 1 out of 100 LGRBs are oriented in our direction this would correspond to a beaming fraction of $f_b$ = 0.01 or a beaming factor of $f_b^{-1}$ = 100.  We will also adopt a uniform cosmology of $\Omega_M$ = 0.23, $\Omega_\Lambda$ = 0.73, \& H$_\circ$ = 71.  As \cite{hb} assume a modified \cite{SalpeterIMF} A IMF \citep{Baldry2003} we will also adopt this IMF.

Given the connection between GRBs and SNe, the first step is to derive the fraction, $f_{cc/SF}$, of core-collapse SNe created per M$_\sun$ of star-formation.  This is estimated (for the \citealt{SalpeterIMF} IMF) as 0.0070 in \cite{SNR007} but without an error given.  This, applied to our R$_{SF}$(z=0) value, gives us a predicted core-collapse SNe rate, $R_{cc}$, of 1.2 $\pm$ 0.2 $\times$ 10$^5$ yr$^{-1}$ Gpc$^{-3}$.  Along with its near corollary the Type II SNe rate, the cc SNe has been well studied and a number of estimates exist in the literature.  \cite{Horiuchi} identified a systematic factor of 2 mismatch between a higher SN rate as estimated from the SFR and the lower directly observed SN rate.  Using a conversion factor of 1 supernova per 114 M$_\sun$ of star-formation (as observed in smaller host galaxies) \cite{Horiuchi} estimates a local cc SNe rate of 1.4 $\times$ 10$^5$ yr$^{-1}$ Gpc$^{-3}$ based on the SFR of the local universe.  This agrees well with the Type II SNe rate of 1.1 $\times$ 10$^5$ from \cite{Lien_math_hell} using similar methodology, but not with the observed cc SNe rate of 7.05$^{+1.65}_{-1.49}$ $\times$ 10$^4$ yr$^{-1}$ Gpc$^{-3}$ of \cite{Weidong_SNe_rates} in a nearby sample of relatively large galaxies. \cite{Horiuchi} believes this discrepancy can be explained due the SNe being optically dim, either intrinsically or due to dust.  This is consistent with observations of $\sim$30\% of LGRBs being dark \citep{Perley_dark_frac, Greiner2011} (lacking an observed optical counterpart presumably because of extinction), since we would expect a lower rate of extinction in the smaller metal poor LGRB host population. Thus we believe this factor of $\sim$2 difference in SNe rates is indeed due to extinction and thus accept the higher SFR inferred cc SNe rate as correct.  We assume that the error between our predicted $R_{cc}$ value and the \cite{Horiuchi} estimate is due the error in the $f_{cc/SF}$ and thus adopt a value with error of 0.007 $\pm$ 0.001 for $f_{cc/SF}$.

%[old rewrite] Our analysis will begin with the core-collapse SNe rate ($R_{cc}$).  
%In section \ref{directly_observed_SNe_rate_insufficient} we will show for our analysis the directly observed SNe rate underestimates the number of SNe required in order to produce the observed LGRB rate.  Comparing the SFR or SNe conversion factor values of 0.0088 of \cite{Lien_math_hell} with 0.00914 of \cite{Horiuchi} suggest that this conversion is known to an accuracy of about 4\% assuming that the local star-formation rate of the universe is known to approximately similar precision allows us to estimate $R_{cc}$ with errors at 1.4 $\pm$ 0.1 $\times$ 10$^5$ yr$^{-1}$ Gpc$^{-3}$.

Next we consider the fraction, $f_{Ibc/cc}$, of cc SNe that are Type Ib or Ic which are collectively referred to as Type Ibc.  For these estimates we require a large homogenous sample of well typed SNe devoid of systematic effects or other biases.  Therefore we appropriate the SNe populations of \cite{Kelly} Figure 1 and compare the number of SNe between the different types therein using this as a representative population.  We also assume Poisson statistics and thus assume the square root of the number of events as the error.  This gives $f_{Ibc/cc}$ values of 0.203 $\pm$ 0.031 and 0.228 $\pm$ 0.027 for galaxy-impartial and galaxy-targeted survey populations respectively.  As these agree within errors we 
%assume this distinction is insignificant for these specific populations and 
adopt the combined value of 0.218 $\pm$ 0.021 for  $f_{Ibc/cc}$ and assume that any difference between the survey populations will be inherently most significant between Type I and II SNe given their difference in brightness.  Notably \cite{Weidong_SNe_rates} gives a direct estimate of the Type Ibc SN rate of 2.58$^{+0.728}_{-0.716}$ $\times$ 10$^4$ yr$^{-1}$ Gpc$^{-3}$ which is actually in rather good agreement with the $\sim$3 $\times$ 10$^4$ yr$^{-1}$ Gpc$^{-3}$ rate we estimate using our $R_{SF}$, $f_{cc/SF}$, and $f_{Ibc/cc}$ values.  This is in contrast to the factor of 2 disagreement between the cc SNe rate derived from observations and from the CSFR. Type I SNe are typically about a magnitude brighter than Type II SNe, which reduces the selection effects introduced by e.g. dust, and the observed and modeled Type I SNe rate are thus in closer agreement.

%Given that Type I SNe are typically about a magnitude brighter than Type II SNe this better agreement between the observed and expected Type Ibc SNe rate compared with the cc SNe rate is not particularly surprising assuming the difference between the measured and SFR inferred SNe rates is indeed due to dust.
 
Next we consider the fraction, $f_{Ic/Ibc}$, of Type Ibc events that are Type Ic.  Using the combined galaxy-impartial and galaxy-targeted \cite{Kelly} populations gives a value of 0.69 $\pm$ 0.09 for $f_{Ic/Ibc}$.  We consider the $f_{Ibc/cc}$ and $f_{Ic/Ibc}$ fractions separately due to the additional difficulty in separating Ib and Ic types which involves determining the presence or absence  of the 5876 {\AA} He I line.  This process is usually omitted from searches interested in identifying Type Ia events and also usually requires later time spectroscopy than is needed to identify the presence or absence of the hydrogen and silicon features used to distinguish between Type I or II and Ia or Ibc SNe respectively. Additionally, by considering the combined fraction of Type Ibc events, it is possible to use Type II SNe rate estimates to determine the core-collapse SNe rate via $R_{II}$ = (1-$f_{Ibc/cc}$) $R_{cc}$.

Next we consider the fraction, $f_{\operatorname{Ic-bl/Ic}}$, of Type Ic events that have relativistic ejecta causing a noticeable Doppler broadening of their spectral features.  Such broad-line features seem to be only associated with Type Ic SNe and are thought to be associated (like LGRBs) with central engine activity \citep{Iwamoto1998}.  However \cite{stats_paper} showed that broad-line Type Ic SNe not only occur primarily at high metallically \citep{Modjaz2008} but follow the metallically distribution of star-formation in the general galaxy population, unlike LGRBs which occur much more frequently in low metallically environments \citep{Fruchter, LeFloch2006, Stanek2007, Wolf, Modjaz2008, Levesque051022, Levesque2, stats_paper, perley_cutoff}.  Again we turn to the sample of \cite{Kelly} which gives a value of 0.21 $\pm$ 0.05 for $f_{\operatorname{Ic-bl/Ic}}$.  As we are using the \cite{Kelly} combined SNe survey population for all of our SNe fractional estimates, we can also directly estimate the fraction of core-collapse SNe that are broad-line Type Ic SNe events, $f_{\operatorname{Ic-bl/cc}}$, without considering the intermediate steps giving a value of 0.032 $\pm$ 0.007.  However these intermediate steps likely contain information on the physical processes that give rise to LGRBs, so we will retain them in our analysis.  Also the individual steps can be more easily compared to current and future SNe surveys than just using the $f_{\operatorname{Ic-bl/cc}}$ value, and consequently improved upon.

%Fruchter1999, Fruchter, LeFlochblue, LeFlochblue2002, Christensen,  

Next we must represent the preference of LGRBs for low-metallicity environments \citep{Fruchter, LeFloch2006, Stanek2007, Wolf, Modjaz2008, Levesque051022, Levesque2, stats_paper, perley_cutoff}.  However there are some LGRBs that do at least appear to occur in high metallically environments \citep{conference_proceedings, Levesque020819B, stats_paper, obs_paper}.  In \cite{diff_rate_letter} we find there to be a sharp cut-off at log(O/H)+12 $\sim$ 8.3 above which the GRB formation rate drops by about a factor of $\sim$30.  Even if this is actually not the case and the actual LGRB environment is always low metallically, we would still need to account for the LGRBs that occur in what we measure to be high metallically host galaxies.  Fortunately this distinction is not only immaterial to this analysis but suggests how we can simply accommodate these exceptions into our fractional methodology. We treat the high metallically LGRBs as if they were also due to additional low metallically star-formation and adjust the fraction accordingly.  Here we represent the fraction of cosmic star-formation at low metallically (log[O/H]+12 $<$ 8.4 in the \citealt{KobulnickyKewley}, henceforth KK04, scale) as $Z^-$, and its corollary the fraction at high metallically star-formation as $Z^+$ (such that $Z^-$ + $Z^+$ = 1).  Thus we calculate the environmental fraction, $f_{Z}$, of adjusted low metallically star-formation by adding to the observed fraction of low metallically star-formation, 
%$f_{LM SF}$
$Z^-$ to, the fraction of high metallically star-formation divided by the relative rate factor, $R_{RF}$. The relative rate factor is the rate (per unit star-formation) at which we produce LGRBs in low metallically environments divided by the rate at which we produce LGRBs in high metallically environments.  We calculated the $R_{RF}$ to be at least 25 and most likely greater than 30 in \cite{diff_rate_letter}.  This yields $f_{Z}$ = \label{f_Z_defined}
%$f_{LM SF} + (1-f_{LM SF})/R_{RF}$
$Z^- + Z^+ / R_{RF}$, which, assuming the value of $Z^-$ = 0.1$_{-0.05}^{+0.1}$ that we estimated from an observational comparison of the distributions LGRB and star-formation with metallically in \cite{stats_paper}, and the $R_{RF}$ value determined in \cite{diff_rate_letter}, gives an adjusted value of $f_{Z}$ = 0.13 $_{-0.05}^{+0.1}$ for the local universe.  The $f_{Z}$ can be thought of as the fraction of low metallically star-formation required if all LGRBs are being formed in low metallically environments.

% , the adjusted environmental fraction of broad-line Type Ic  events that are in low metallicity environments (which we assume to be the fraction of star-formation at low metallicity) summed with the fraction at high metallicity divided by our metallicity relative rate factor ($f_{LM SF} + [1-f_{LM SF}]/R_{RF}$),
 
Give that not all broad-line Type Ic SNe, even those in low metallically environments, necessarily generate an LGRB, we add an efficiency fraction, $f_{eff}$, to represent the fraction that do, and which effectively contains our lack of understanding on what precise progenitor conditions result in an LGRB.  Given that we do not fully understand the exact progenitor properties that lead to a broad-line Type Ic SNe being accompanied by an LGRB, estimating this efficiency fraction is the principal motivation of this paper.

%Finally we add an efficiency fraction, $f_{eff}$, to represent the fraction of broad-line Type Ic SNe events occurring in low metallicity environments which proceed to form LGRBs.

%Give that we donÕt fully understand the exact progenitor properties that lead to a broad-line Type Ic SN being accompanied by a LGRB, we include an additional efficiency factor

 %As we have not idea what this rate is we will begin by assuming that all such suitable SNe events form LGRBs.  Later we will solve for this fraction based on the observed LGRB rate and thus in practice this fraction will also contain any errors in the other estimates.
 
ÒIn order to consider only LGRBs aligned in our direction we must further consider the beaming fraction, $f_{b}$ given by the burst's beaming angle.  There are two wildly different values for $f_{b}^{-1}$ of 520 $\pm$ 85 and 75 $\pm$ 25 from \cite{Frail} and \cite{Guetta} respectively.  The different values are calculated via different methodologies.  The primary difference being that the \cite{Guetta} method assumes the existence of a large number of low-luminosity bursts which cannot be observed except at the closest redshifts (see \citealt{Piran_rome_proceeding} for a more detailed yet still brief comparison of the two methodologies).  We also note the consistency between the beaming angle of the \cite{Frail} population (see their Figure 1) with that of the {\it Swift} population (see \citealt{Ryan} Figure 4).  However the \cite{Frail} methodology makes certain simplifications in their assumed jet structure and assumes any optical break to be a jet break, which is now known to be a poor approximation.  \cite{Racusin2009} analyzed the X-ray afterglows of two years of {\it Swift} GRBs ($\sim$250 objects) using the relation between the spectral and light curve slopes to determine which breaks in the X-ray light curves were likely jet breaks, and in these cases derived the jet opening angle.  One issue is that to determine the jet angle, one of the parameters that is required is the GRB isotropic-equivalent energy release, E$_{iso}$. This requires the burst redshift to be known and thus is not available for most LGRBs. Even in those cases where E$_{iso}$ is known, the range in opening angles is fairly large, and the sample sizes appear insufficient to confidently determine a distribution, and thus the mean angle.  For example \cite{Gao2010} looked at the jet opening angles for a sample of 28 long GRBs, and they found the mean and standard deviation to be 3.42 $\pm$ 9.07 degrees!

\begin{figure*}[!t]
\begin{equation}
\label{LGRB_drake}
%R_{aLGRB} = R_{SF} \times f_{cc/SF} \times f_{Ibc/cc}  \times f_{Ic/Ibc}  \times f_{\operatorname{Ic-bl/Ic}} \times (f_{LM SF} + {1-f_{LM SF} \over R_{RF}})  \times f_{eff}  \times f_{b}
R_{aLGRB} = R_{SF} \times f_{cc/SF} \times f_{Ibc/cc}  \times f_{Ic/Ibc}  \times f_{\operatorname{Ic-bl/Ic}} \times \bigg( Z^- + {{Z^+} \over R_{RF}}\bigg)  \times f_{eff}  \times f_{b}
\end{equation}
%\vspace*{-0.5 cm}
\end{figure*}

This process yields equation \ref{LGRB_drake}.  Since there is considerable uncertainty in the beaming fraction and the efficiency fraction is unknown we will combine the opening angle and efficiency into a single unknown parameter, the Combined efficiency and beaming fraction, $f_{ceb}$, so that we can provide estimates that will allow the reader to easily estimate the efficiency fraction with their choice of beaming fraction (i.e. $f_{ceb}$ = $f_{eff}$  $\times$ $f_{b}$).  This along with a simplification of the SNe fractions yields equation \ref{LGRB_drake_simp}.

\begin{figure*}[!t]
\begin{equation}
\label{LGRB_drake_simp}
%R_{aLGRB} = R_{SF} \times f_{cc/SF} \times f_{Ibc/cc}  \times f_{Ic/Ibc}  \times f_{\operatorname{Ic-bl/Ic}} \times (f_{LM SF} + {1-f_{LM SF} \over R_{RF}})  \times f_{eff}  \times f_{b}
R_{aLGRB} = R_{SF} \times f_{cc/SF} \times f_{\operatorname{Ic-bl/cc}} \times \bigg( Z^- + {{Z^+} \over R_{RF}}\bigg)  \times f_{ceb}
\end{equation}
%\vspace*{-0.5 cm}
\end{figure*}

\begin{table*}
\begin{center}
% \color{white}
\caption{Table of values for equation \ref{LGRB_drake}.    \label{LGRB_drake_values}}
\vspace{-0.1 cm}
\begin{tabular}{ccccccc}
\hline
\hline
Variable & Description & Value & Source \\
\hline
$f_{cc/SF}$ & Number of cc SNe per M$_\sun$ of SF & 0.007 $\pm$ 0.001 & \cite{SNR007} \\
$R_{cc}$    & Local rate of core collapse SNe$^g$ & 1.4 $\pm$ 0.1 $\times$10$^5$$^{i}$ & \cite{Horiuchi} \\
%$N_{cc}$    & Number of core collapse SNe$^b$ & $\sim$10$^8$$^e$ & \cite{Lien_math_hell} \\
%$f_{Ibc/cc}$   &  Fraction of core collapse that are Ibc's   & \specialcell{$\sim$0.25 \\0.22$^{d,g}$ $\pm$ 0.02$^a$}  &   \specialcell{\cite{Guetta2007} \\ \cite{Kelly}} \\
$f_{Ibc/cc}$   &  Fraction of core collapse SNe that are Ibc's   & 0.22$^{d}$ $\pm$ 0.02$^a$ & \cite{Kelly} \\
$f_{Ic/Ibc}$   &  Fraction of Ibc's that are Ic's &  0.69$^d$ $\pm$ 0.09$^a$  &  \cite{Kelly}   \\
$f_{\operatorname{Ic-bl/Ic}}$   &  Fraction of Ic's that are broad-line & 0.21$^d$ $\pm$ 0.05$^a$  &  \cite{Kelly}  \\
%$f_{LM SF / SF}$
$Z^+ $  & Low metallicity star-formation fraction & 0.1$_{-0.05}^{+0.1}$$^c$   &  \cite{stats_paper}  \\
$R_{RF}$   & Metallicity relative rate factor  & $\sim$30 or $>$25  & \cite{diff_rate_letter}  \\
$f_{eff}$   & Efficiency fraction  & Unknown  &    \\
$f_{b}$   & Beaming fraction  & \specialcell{0.0019 $\pm$ 0.0003 \\ 0.013 $\pm$ 0.004}  &  \specialcell{\cite{Frail} \\ \cite{Guetta}}  \\
%$R_{aLGRB}$ & Local rate of aligned LGRBs$^g$ &  \specialcell{0.42$^{+0.9}_{-0.4}$ \\ $\sim$0.825$^{h,j}$} & \specialcell{\cite{Lien} \\ \cite{Guetta2007}} \\
$R_{aLGRB}$ & Local rate of aligned LGRBs$^g$ &  0.42$^{+0.9}_{-0.4}$ & \cite{Lien_err}  \\
$N_{aLGRB}$ & Total number of aligned LGRBs$^b$ & 4568$^{+825}_{-1429}$$^e$ & \cite{Lien_err} \\
\noalign{\vskip 0.5mm} 
\hline
\noalign{\vskip 0.5mm} 
$Z^+ $ = $1 - Z^-$   & High metallicity star-formation fraction & $0.9_{-0.05}^{+0.1}$$^c$   & Calculated from above   \\
$f_{Z}$ = $Z^- + Z^+ / R_{RF}$ &  Adjusted environmental fraction & $0.13_{-0.05}^{+0.1}$   &  Calculated from above   \\
$f_{\operatorname{Ic-bl/Ibc}}$ = $f_{Ic/Ibc} \times f_{\operatorname{Ic-bl/Ic}}$ & Fraction of Ibc's that are broad-line Ic's & \specialcell{$\approx$0.07 \\  0.14$^{e}$ $\pm$ 0.04} & \specialcell{\cite{Guetta} \\ Calculated from above} \\
$f_{\operatorname{Ic-bl/cc}}$ & Fraction of core collapse SNe that are Ic-bl & 0.032 $\pm$ 0.007 & Calculated from above   \\
$f_{ceb}$ = $f_{eff}  \times f_b$ & Combined efficiency and beaming fraction & 4000 $\pm$ 2000 & this paper - section \ref{ceb_value} \\
\noalign{\vskip 0.5mm} 
\hline
\end{tabular}
\end{center}
%\caption
\vspace{-0.2 cm}
%{Table of values for equation \ref{LGRB_drake}.} \\
$^a$ Assumed to be Poisson therefore standard deviation is estimated as the square root of the counts. \\
$^b$ Across all redshifts. \\
$^c$ Value for redshift zero only. \\
$^d$ Calculated from totals across both plots in \cite{Kelly} Figure 1.  These plots are assumed to be a representative population. \\
%$^e$ Composite of above shown for comparison. \\
$^e$ Value for entire universe. \\
$^f$ The lack of undiscriminated Type Ibc's does pose the question if these events are under represented. \\
$^g$ Value for local universe per GPc per year. \\
%$^h$ Matches \cite{Lien_math_hell} Type II SNe rate divided by (1 - $f_{Ibc/cc}$). \\
%$^j$ Computed based on $\frac{3}{4}$ of BATSE GRBs being LGRBs. \\
\end{table*}

\section{The Cosmic LGRB and SFR Rates} \label{rate_comp}

Altogether these steps produce equations \ref{LGRB_drake} and \ref{LGRB_drake_simp}, which gives an estimate of the rate of aligned LGRBs and has a notable similarity to the Drake equation for estimating the number of inhabited extrasolar planets in our galaxy.  Unlike the Drake equation however this equation has the considerable advantage of the LGRB rate, R$_{aLGRB}$, now being reasonably well known thanks to the seminal work of \cite{Lien} (revised in \citealt{Lien_err} and expanded in \citealt{Graff}).  Their detailed analysis and simulation of the {\it Swift} BAT detector gives a rate of 0.42 yr$^{-1}$ Gpc$^{-3}$ for aligned LGRB events in the local universe \citep{Lien_err}. Unfortunately as we will address in section \ref{local_problem} the parametric forms of the local LGRB and CSFR are sufficiently different to preclude our believing this ``local" rate is truly representative at z $<$ 1.  Nevertheless, \cite{Lien}, and in a comparable but updated analysis, \cite{Graff}, provide reasonable observational estimates on the LGRB rate as a function of redshift (at least in the 1 $<$ z $<$ 4 range).  

%Still \cite{Lien} and an expanded version of this methodology allowing for a more observationally based fitting of the LGRB rate, \cite{Graff}, provide reasonable observational estimates on the LGRB rate as a function of redshift.  

%This analysis can also be extended to the universe at large by replacing the rates per unit volume of core collapse SNe ($R_{cc}$) and aligned LGRBs ($R_{aLGRB}$) with their respective number of observable events per year across the entire universe (which we will refer to as $N_{cc}$ and $N_{aLGRB}$).  However the fraction of low metallicity star-formation also changes with redshift, ignoring this gives us crude upper and lower limits on the number of LGRBs occurring in the observable universe per year by assuming that all star-formation is at low metallicity or that the low metallicity star-formation fraction in the local universe remains constant respectively. This gives respective upper and lower bound values of ** and ** which enclose the estimated value of ** from \cite{Lien_err}.

\subsection{The Evolution of Metallicity} \label{Metallicity_Evolution}

%Since we can not directly compare the results of equation \ref{LGRB_drake} at z = 0 and the fraction of low metallicity star-formation changes with redshift we must at least approximate it.

Understanding the evolution of the low metallicity star-formation fraction 
%($f_{LM SF / SF}$)
($Z^-$) with redshift is complicated and a task we hope to address in upcoming work.  While there are some existing observational \citep{Tiantian2013, Zahid2013} and theoretical \citep{Yates2012} efforts to understand metallicity evolution with redshift, none of them enable us to estimate the $Z^-$ term as a function of redshift.  What we need is a census of the distribution of star-formation below a set metallicity cutoff throughout the universe as a function of redshift.  While we hope to address this with simulations and perhaps observations in a future work, for now however we can place some constraints with our existing knowledge and will attempt some estimates based on first principles.  Furthermore by trying and then evaluating a diverse range of possibilities we may be able to shed some insight on the evolution of metallicity itself.

As determined in \cite{stats_paper}, we know that the Z$^+_{now}$ $\approx$ 0.9 in the local universe and we can assume that  Z$^+_\circ$ = 0 primordially.  Furthermore as metal production is the result of star-formation we can assume that the rise in Z$^+$ should correspond to the integral of the Cosmic Star-Formation Rate (CSFR).  The remaining question is what other factors will effect the relationship between metallicity enhancement and star-formation.  Ideally we would consider not only the creation of metals (i.e. star-formation) but also account for the removal and dilution of metals by outflows and inflows.   Given the uncertainties in the effect of outflows and inflows, and that these are most likely second order effects, we only consider the effect on metallicity evolution from $Z^+$ and z. 
%Here however we are limited as the only independent variables we have easily available are are $Z^-$ and z (and their dependent variables such as $Z^+$ and the cosmological parameters like age respectively).  
Thus we consider only the effects of the scale factor and of the current metallicity on the efficiency of star-formation in increasing the fraction of high metallicity star-formation (i.e. log(O/H)+12 $>$ 8.3 in KK04 scale).  For the scale factor we consider that the efficiency of star-formation in increasing the metal content of the universe might increase by a factor of (1+z) or (1+z)$^3$ (and also that scale factor might play no role in metal enrichment efficiency).  The factor of (1+z) takes into account the compression of the Universe, and the factor of (1+z)$^3$ corresponds to the increase in the specific SFR, both of which we can reasonably imagine would effect the evolution of Z$^-$.
%Given that the Specific Star-Formation Rate is thought to increase by approximately the latter factor this is not an unreasonable postulation nor is the thought that the compression of the universe might effect this relationship.
It is important to factor in that once gas enrichment reaches a certain metallicity, the star formation at low metallicity will no longer increase, at least if we assume that mixing is inefficient.  Thus the effect of SFR on the increase of Z$^+$ would effectively be reduced by a factor of Z$^+$.
%  To consider the impact of the material already at high metallicity on the relationship between metallicity enhancement and star-formation we consider that perhaps if mixing is poor then once gas has been enriched above the  metallicity cut off further enrichment of (via star-formation with) the Z$^+$ gas is would not effect Z$^-$ hence reducing the effect of SFR on enrichment by a factor of Z$^+$. 
 I.e. when at some point a third of the universe has been enriched to high metallicity we assume that star-formation will only be two-thirds as effective at metal enrichment as when Z$^+$ = 0.  Of course if the universe is well mixed than super enriched gas would still contribute as it was diluted by metal poor gas.  This gives us six relationships to test: CSFR, CSFR$\times$(1+z), CSFR$\times$(1+z)$^3$, CSFR$\times$Z$^-$(z), CSFR$\times$(1+z)$\times$Z$^-$(z), CSFR$\times$(1+z)$^3$$\times$Z$^-$(z),

%\begin{equation}
%\begin{split}
%\label{Z_eqn}
% &\quad  Z^+_{n+1} = Z^+_n + a \times CSFR_n \\
% &\quad  Z^+_{n+1} = Z^+_n + a \times CSFR_n\times(1+z_n) \\
% &\quad  Z^+_{n+1} = Z^+_n + a \times CSFR_n\times(1+z_n)^3 \\
% &\quad  Z^+_{n+1} = Z^+_n + a \times CSFR_n\times Z^-_n \\
% &\quad  Z^+_{n+1} = Z^+_n + a \times CSFR_n\times(1+z_n)\times Z^-_n \\
% &\quad  Z^+_{n+1} = Z^+_n + a \times CSFR_n\times(1+z_n)^3\times Z^-_n
%\end{split}
%\end{equation}

%\begin{figure*}[!!]
\vspace*{-0.3 cm}
\begin{equation}
\begin{tabular}{lc} \label{Z_eqn} \\
$Z^+_{n+1} = Z^+_n + a \times CSFR_n$ & (purple) \\
$Z^+_{n+1} = Z^+_n + a \times CSFR_n\times(1+z_n)$ & (blue) \\
$Z^+_{n+1} = Z^+_n + a \times CSFR_n\times(1+z_n)^3$ & (cyan) \\
$Z^+_{n+1} = Z^+_n + a \times CSFR_n\times Z^-_n$ & (yellow) \\
$Z^+_{n+1} = Z^+_n + a \times CSFR_n\times(1+z_n)\times Z^-_n$ & (orange) \\
$Z^+_{n+1} = Z^+_n + a \times CSFR_n\times(1+z_n)^3\times Z^-_n$ & (red) \\
\end{tabular}
\end{equation}
%\end{figure*}

\begin{figure}[h]
\begin{center}
%the h! tells latex you want the Figure inserted here.
\includegraphics[width=.48\textwidth]{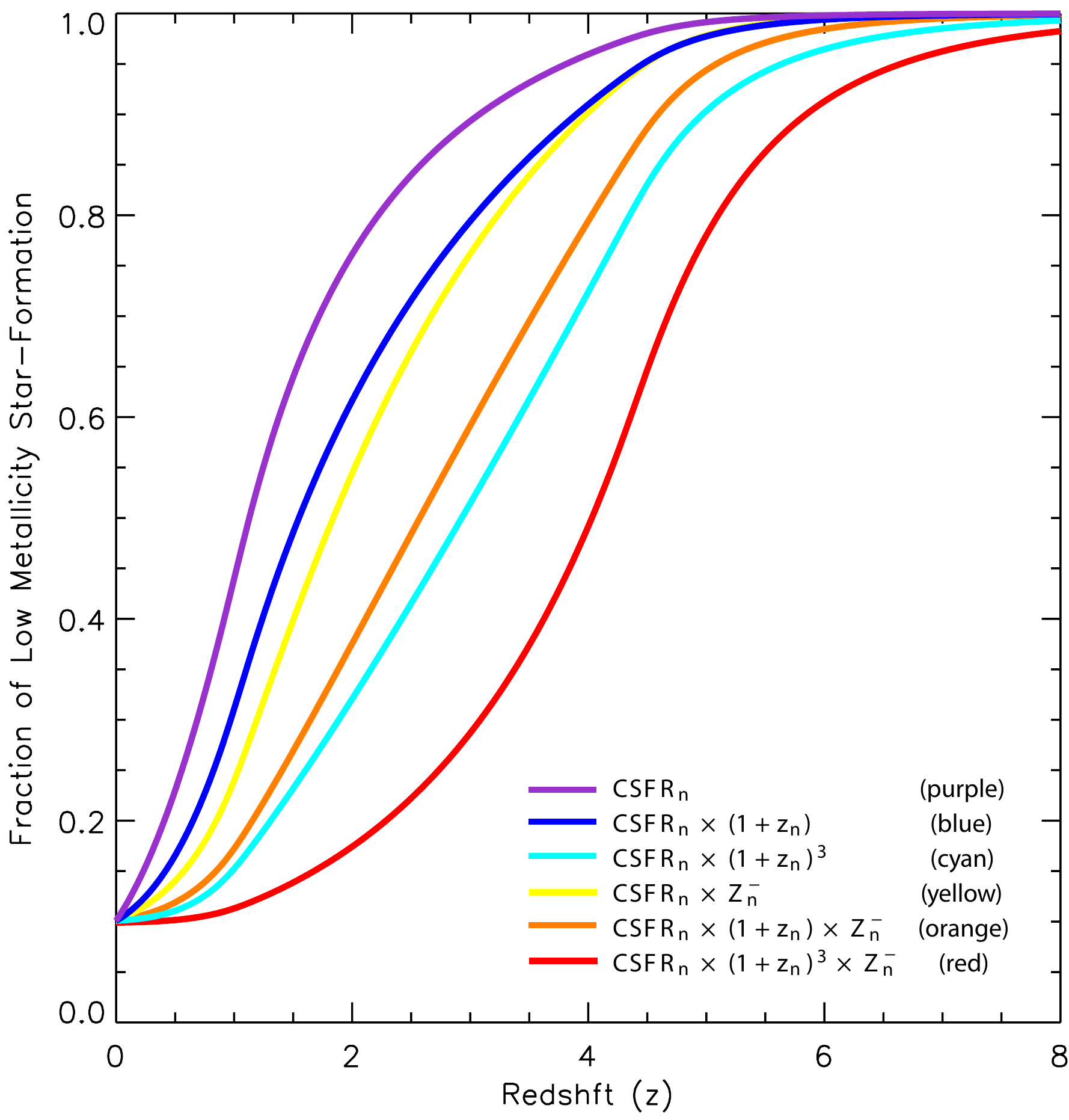}
\caption{\label{Zp_z_evolution_cases_plot} Fraction of low metallicity (log(O/H)+12 $<$ 8.4 in the KK04 scale) as a function of redshift, $Z^-(z)$, for the different cases of metallicity evolution determined by iterating the cases given in equation \ref{Z_eqn} such that  $Z^+$(z=0) =  0.9 and $Z^+$(z=20) =  0.  We believe these 6 cases provide a reasonable distribution of possibilities for the evolution of metallicity with redshift.}
\end{center}
\end{figure}

We thus iterate these relationships in the form of equation \ref{Z_eqn} in million year increments from z=20 to now (assuming Z$^+_{z=20}$ = 0) with the constant $a$ selected (for each relationship) such that Z$^+_{now}$ = 0.9 $\pm$ 0.001 with the results shown in Figure \ref{Zp_z_evolution_cases_plot} (this error represents only the tolerance used in determining the $a$ values).  As each of these relationships form a distinct metallicity evolution test case we find it useful to assign a name to each case --- for simplicity and lack of a short physically descriptive names we assign names to the metallicity evolution cases corresponding to the colors we use to plot them in the Figures.

\subsection{Difficulties in estimating the local LGRB Rate} \label{local_problem}

\begin{figure}[h]
\begin{center}
%the h! tells latex you want the Figure inserted here.
\includegraphics[width=.48\textwidth]{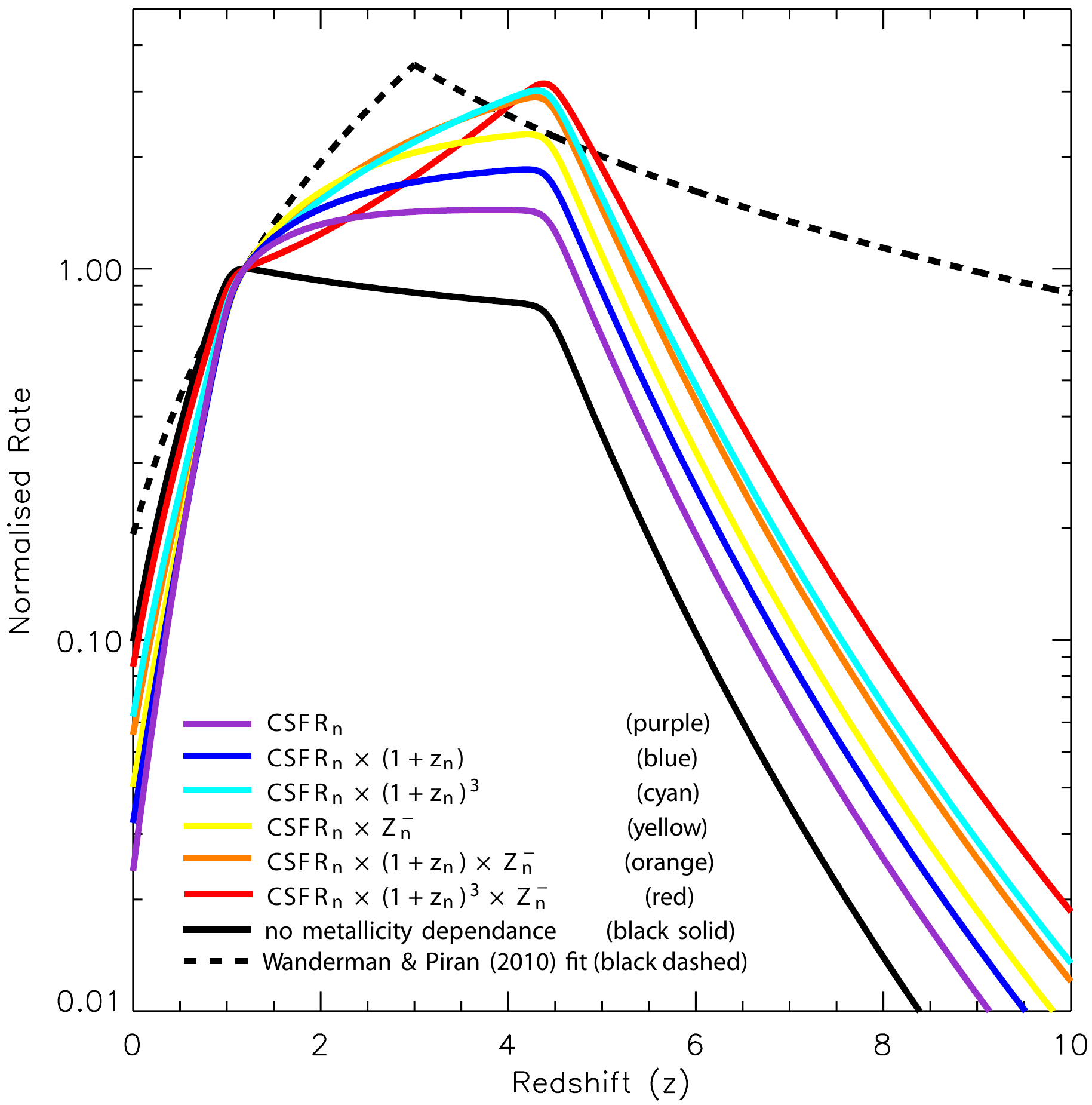}
\caption{\label{Normalised_rate_plot} CSFR at low metallicity (colored lines) as a function of redshift compared with the total CSFR (black solid line) and the LGRB rate estimates of \cite{wp} (dashed solid line) with all values normalized at the redshift of the CSFR peak  (z $\approx$ 1.2).  The CSFR at low metallicity is determined as the product of the CSFR and our low metallicity fraction estimates, $Z^-(z)$, from Equation \ref{Z_eqn} as shown in Figure \ref{Zp_z_evolution_cases_plot}. }
\end{center}
\end{figure}

Broken power law estimates of the CSFR(z) give an initial evolution of (1+z)$^x$ with x being about 3.4 \citep{hb} whereas estimates of the LGRB rate (typically fit from now out to z of 3 or 4) go with x being about 2.1 \citep{wp, Lien}.  As a result of the small number of LGRBs at z $<$ 1, any functional form fitted to the LGRB is very poorly constrained in this range. A simple power law is commonly assumed, where the LGRB rate at 1 $<$ z $<$ 3 weigh more heavily.  While one might expect metallicity evolution to be able to explain the CSFR vs.~LGRB rate slope difference metallicity evolution actually acts in the opposite direction (assuming of course that the metal content of the universe can only increase with time).  Assuming that the LGRB rate must follow the CSFR (with some enhancement at low metallicities), this suggests that the fitted z = 0 LGRB rate is over estimated by at least a factor of 2 when the LGRB rate is estimated as a continuous power law out to z $\sim$ 3.  This discrepancy between the observed LGRB rate extrapolated down to z = 0 and the lower LGRB expected from the CSFR is not particularly observable because, due to LGRBs being relatively rare events in general, the nearby (non sub-luminous) LGRB rate is low, which means that the statistical uncertainty on the observed local rate is large.  For example there are only 5 LGRBs (with E$_{iso}$ $>$ 10$^{50}$ ergs) at z $<$ 0.3.  Also, as shown in Figure \ref{Normalised_rate_plot}, while the CSFR is only 50\% of the normalized LGRB rate estimates of \cite{wp}, by redshift z = 0.3 this difference is only 70\%, at at z = 0.5 it si almost negligible.  There is therefore little constraint to the functional form fitted to the LGRB rate at z $\lesssim$ 1 and thus the functional form adopted is usually an single power-law fit in the 0 $<$ z $\lesssim$ 3 range. By contrast the CSFR(z) seems to undergo a radical $\sim$3.5 difference in power-law slope at z $\approx$ 1.  Since there are far more LGRBs in the 1 $<$ z $<$ 3 range than the 0 $<$ z $<$ 1 range the tendency is to compare and normalize the LGRB rate and the CSFR in this range.

In Figure \ref{Normalised_rate_plot} we show the (1+z)$^{2.1}$ LGRB rate of \cite{wp} along with the CSFR(z) of \cite{hb}.  In addition, we show the CSFR(z) convolved with our 6 low metallicity evolution functions described in section \ref{Metallicity_Evolution} (see Figure \ref{Zp_z_evolution_cases_plot}). All plotted curves are normalized to unity at the peak redshift of the CSFR (z $\approx$ 1.2).  All the CSFR based LGRB estimates are thus at least a factor of two lower than the 0 $<$ z $<$ 3 power-law fit of \cite{hb}.  Given the limited number of LGRBs in the 0 $<$ z $<$ 1 range we believe that the LGRB rate does indeed drop as the CSFR based estimates predict, but as the LGRB rate is relatively unconstrained in this region usually the simplest model is fitted.  Furthermore estimates of the z = 0 LGRB rate (c.f. \citealt{wp, Lien, Graff}; etc.) that do not take this into account (as \citealt{butler} did) will be systemically over estimated.% it due to their inherently being estimates of the LGRB rate at higher redshifts extrapolated back to z = 0 using this inaccurate functional form of a consistent 0 $<$ z $<$ 3 power-law fit.

To avoid falling victim to this trap ourselves we will not attempt to compare the LGRB and CSFR rates at z = 0 despite the local universe being the only region where we have observational information on the distribution of star-formation with metallicity.  Instead we will use our metallicity evolution models from section \ref{Metallicity_Evolution} to preform this analysis at z $>$ 1 where both the CSFR(z) and the LGRB rate is well known.

%match our well constrained 

% 0 $<$ z $\lesssim$ 3 power-law

% Although the comparison to the CSFR would suggest that the simple power law fitted to the LGRB rate out to z~3 is incorrect, and that an additional break is likely present at z<3

%given the steep decline of the LGRB rate and the few (non sub-luminous) LGRB events at such low redshifts.  For example there are only 5 LGRBs (with E$_{iso}$ $>$ 10$^{50}$ ergs) at z $<$ 0.3 and, as shown in Figure \ref{Normalised_rate_plot}, at z = 0.3 the normalized total CSFR is already 70\% of the normalized LGRB rate estimates of \cite{wp}.

%Understanding the evolution of the Low Metallicity Star-Formation Fraction ($f_{LM SF / SF}$) with redshift is complicated and a task we hope to address in upcoming work.  However we can place some constraints with our existing knowledge.   \cite{stats_paper} looks at the metallicity of LGRB hosts and galaxies out to z $\sim$ 1 and finds that at highest of these redshifts the brightest of these galaxies are still about 0.4 dex higher than the typical  LGRB metallicity whereas the brightest galaxies at low redshift (z $\lesssim$ 0.04) are about 0.8 dex higher than the typical LGRB metallicity.  Furthermore \cite{Greiner_trace} find that GRBs successfully trace UV metrics of cosmic star formation over the range 3 $<$ z $<$ 5 implying that there is no longer a dominant metallicity effect acting on this population. Therefore we adopt a crude model of assuming that the local $f_{LM SF / SF}$ value holds out to z $<$ 2 and that all star-formation is at low metallicity (i.e. $f_{LM SF / SF}$ = 1) for z $>$ 2.

\subsection{LGRB rate and Low Metallicity CSFR Redshift Evolution}

Having estimates for the cosmic low metallicity star-formation rate, the next step is to compare them with the LGRB rate as a function of redshift.  The latter has been estimated in \cite{wp} and \cite{butler}.  Here we will primarily use the \cite{wp} data for comparison as its methodology of binning LGRBs by redshift is more easily reproducible and comparable to our methodology.  We restrict our fitting to the redshifts below 4 to avoid being significantly affected by the slope of the high redshift term in the \cite{hb} piecewise power-law CSFR estimate, which becomes highly uncertain at z $>$ 4.  The approach is almost identical to that shown in Figure \ref{Normalised_rate_plot} except that instead of normalizing our values at a common redshift here we will normalize each metallicity evolution case such that it provides the lowest least squares fit with the \cite{wp} data.

\begin{figure}[h!]
\begin{center}
%the h! tells latex you want the Figure inserted here.
\includegraphics[width=.48\textwidth]{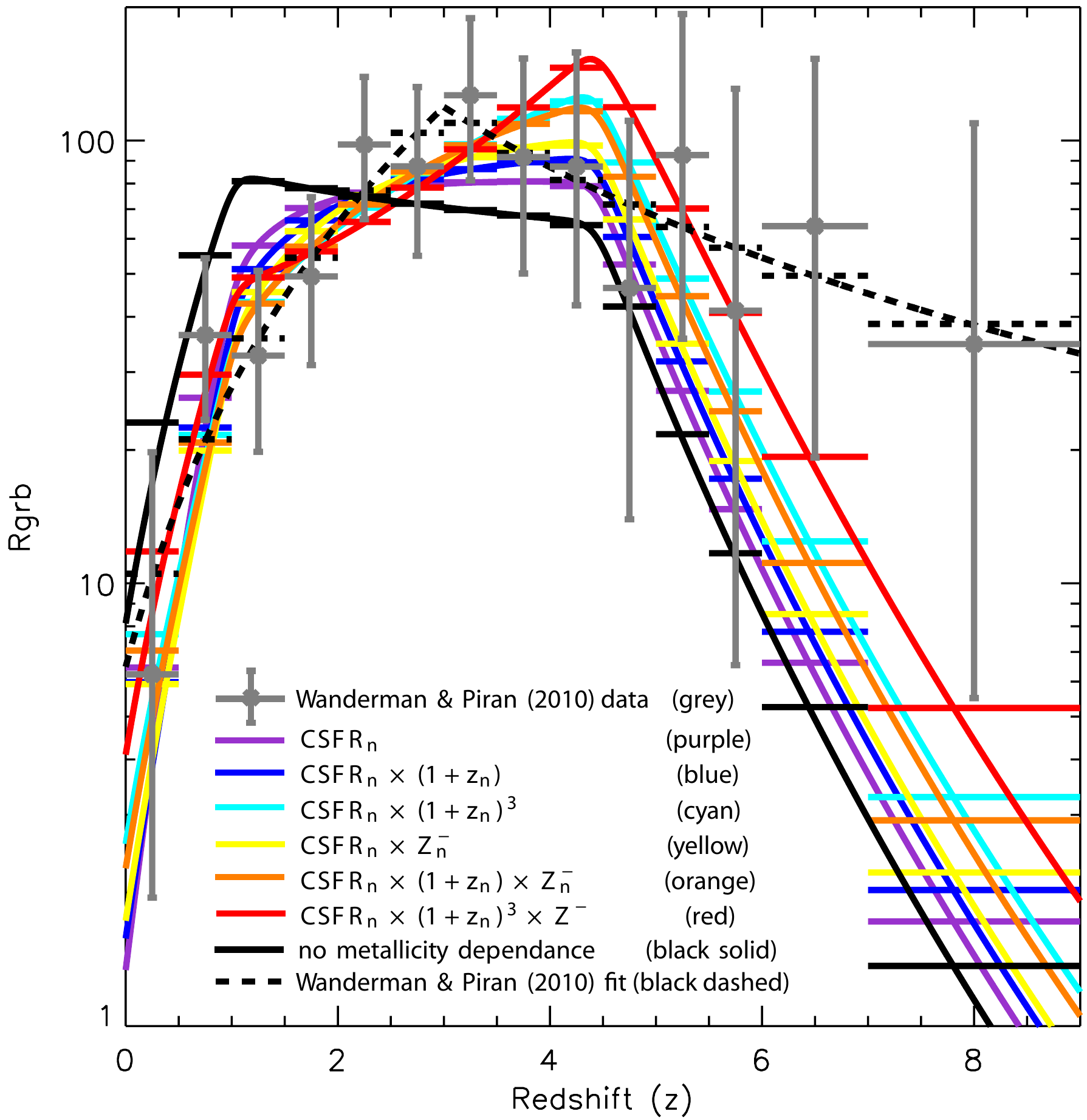}
\caption{\label{Fitted_rate_plot} Here we repeat the plotting in Figure \ref{Normalised_rate_plot} except now normalize all lines to the logarithmic best fit in the 0 $<$ z $<$ 4 range of the \cite{wp} rate estimates shown in grey.  As this requires binning all the lines into histograms for fitting, additional horizontal bars of matching color indicate the level in each bin.  As the binning process accounts for the change in comoving volume of the universe within individual bins the matching lines and bars do not strictly intersect in the center of the bins.}
\end{center}
\end{figure}

\begin{table*}
\begin{center}
\caption{\label{Fitted_rate_values} Metallicity Evolution Cases Evaluated}
\vspace{-0.1 cm}
\begin{tabular}{cccccccccccc}
\hline
\hline
\multirow{2}{*}{Metallicity Evolution Case} & Plot  & Normalized Error \\
& Color & fit: 0 $<$ z $<$ 4  \\
\hline
\multicolumn{1}{l}{$Z^+_{n+1} = Z^+_n + a \times CSFR_n$} & purple & 1.21  \\
\multicolumn{1}{l}{$Z^+_{n+1} = Z^+_n + a \times CSFR_n\times(1+z_n)$} & blue & 1.02  \\
\multicolumn{1}{l}{$Z^+_{n+1} = Z^+_n + a \times CSFR_n\times(1+z_n)^3$} & cyan & 0.88 \\
\multicolumn{1}{l}{$Z^+_{n+1} = Z^+_n + a \times CSFR_n\times Z^-_n$} & yellow & 0.94 \\
\multicolumn{1}{l}{$Z^+_{n+1} = Z^+_n + a \times CSFR_n\times(1+z_n)\times Z^-_n$} & orange & 0.86  \\
\multicolumn{1}{l}{$Z^+_{n+1} = Z^+_n + a \times CSFR_n\times(1+z_n)^3\times Z^-_n$} & red & 1.34 \\

%$Z^+_{n+1} = Z^+_n + a \times CSFR_n$} & purple & 1.21  \\
%$Z^+_{n+1} = Z^+_n + a \times CSFR_n\times(1+z_n)$} & blue & 1.02  \\
%$Z^+_{n+1} = Z^+_n + a \times CSFR_n\times(1+z_n)^3$} & cyan & 0.88 \\
%$Z^+_{n+1} = Z^+_n + a \times CSFR_n\times Z^-_n$} & yellow & 0.94 \\
%$Z^+_{n+1} = Z^+_n + a \times CSFR_n\times(1+z_n)\times Z^-_n$} & orange & 0.86  \\
%$Z^+_{n+1} = Z^+_n + a \times CSFR_n\times(1+z_n)^3\times Z^-_n$} & red & 1.34 \\

$Z^+ = 0$ (i.e. no metallicity dependance) & black solid & 4.48  \\
Original \cite{wp} fit & black dashed & 1  \\
\hline
\end{tabular}
\end{center}
\vspace{-0.2 cm}
%\caption{\label{Fitted_rate_values} 
{Least squares logarithmic errors of the various metallicity evolution cases (colored lines), the cosmic star-formation rate history (black line), and the two piece pow-law fit of \cite{wp} (dashed black line) with respect to the \cite{wp} data (grey points with error bars) as shown in Figure \ref{Fitted_rate_plot}.  This fitting and associated error computation is limited to the 0 $<$ z $<$ 4 range so the slope of the poorly known final third high-redshift term in the \cite{hb} CSFR history does not effect results.} 
\end{table*}

We find that 3 of our metallicity evolution cases provide a better fit to the \cite{wp} data than the two step broken power-law fit given in \cite{wp} (with an additional forth case with comparable error).  This strongly suggests that metallicity evolution can explain the distribution of LGRBs as a product of the CSFR historyÒ, at least out to z $\sim$ 4.  A similar conclusion was suggested in \cite{butler}.  Having established a general consistency in shape between the data and our model across redshift, we will now apply the methodology of section \ref{drake} to numerically compare the LGRB rate and CSFR history.  Essentially the remaining exercise is one of scaling.

\subsection{Comparing the Number of LGRB and Their Progenitors} \label{ceb_value}

While \cite{wp} and \cite{butler} have looked at the distribution of LGRBs with redshift, an alternate approach pioneered in \cite{Lien} is to fully simulate the {\it Swift} Burst Alert Telescope (BAT) performance on a set of synthetic LGRB events varying their number, redshift distribution, and overall parameters so as to reproduce the observed {\it Swift} population.  The large number of {\it Swift} events with measured redshifts allows this technique to be employed on the {\it Swift} LGRB redshift distribution, whereas it would not have been feasible on the much smaller collection of GRBs with known redshifts from all other missions combined.  However the primary advantage of the \cite{Lien} methodology is that it allows not just an assessment of the redshift distribution of LGRBs but also their absolute number.  However, while \cite{Lien} assumes a rather conventional cosmic LGRB rate described by a fixed piecewise power-law with the pieces meeting at z = 3.6, and an initial power-law slope almost identical to that of \cite{wp}, this methodology was expanded in \cite{Graff} to allow much more flexibility in fitting parameters.

Thus we compare the amount of low metallicity star-formation in the universe (as a function of redshift) with the \cite{Lien} and the \cite{Graff} estimates, 
%Thus we compare the absolute amount of cosmic star-formation to the estimated number of LGRBs by 
using equation \ref{LGRB_drake_simp} to estimate the combined efficiency and beaming factor, $f_{ceb}^{-1}$, since all the other parameters are reasonably well known.  We limit this comparison to redshifts above 1 because the LGRB rate is fairly unconstrained at z $<$ 1 due to small number statistics (see section \ref{local_problem}).
%   This would otherwise impacted the results as the LGRB rate is not particularly sensitive to the different much steeper slope of the CSFR at z $<$ 1 and neither \cite{Lien} nor \cite{Graff} used parametric forms allowing for different LGRB rate sloped above and below z = 1 as is present in the CSFR.
As before we also limit this comparison to redshifts below 4 as the slope of the CSFR at higher redshift is poorly known.  This has the added advantage of mostly excluding the slope of the post peak decline in the LGRB rate, which has quite high uncertainties. %, from consideration.  

\begin{table*}
\begin{center}
\caption{\label{ceb_values} $f_{ceb}$ Estimates for Each Metallicity Evolution Case}
\vspace{-0.1 cm}
\begin{tabular}{cccccccccccccc}
\hline
\hline
\multirow{2}{*}{Metallicity Evolution Case} & Plot  & Lien et al. & \multicolumn{3}{c}{\cite{Graff}} \\
& Color & (\citeyear{Lien_err}) & RF & AB & NN \\
\hline
%\multicolumn{1}{|l|}{ \color{gray} $Z^+_{n+1} = Z^+_n + a \times CSFR_n$} & \color{gray} 5786 $\pm$ 2316 &  \color{gray} 7752 $\pm$ 2478 &  \color{gray} 7776 $\pm$ 2449 &  \color{gray} 7462 $\pm$ 2522   \\
%\multicolumn{1}{|l|}{ \color{gray} $Z^+_{n+1} = Z^+_n + a \times CSFR_n\times(1+z_n)$} &  \color{gray} 4771 $\pm$  \color{gray} 1535 &  \color{gray} 6432 $\pm$ 1571 &  \color{gray} 6454 $\pm$ 1547 &  \color{gray} 6184 $\pm$ 1601 \\
%\multicolumn{1}{|l|}{$Z^+_{n+1} = Z^+_n + a \times CSFR_n\times(1+z_n)^3$} & 2818 $\pm$ 532 & 3836 $\pm$ 412 & 3851 $\pm$ 395 & 3687 $\pm$ 466    \\
%\multicolumn{1}{|l|}{ \color{gray} $Z^+_{n+1} = Z^+_n + a \times CSFR_n\times Z^-_n$} &  \color{gray} 4266 $\pm$ 1070 &  \color{gray} 5785 $\pm$ 1036 &  \color{gray} 5807 $\pm$ 1014 &  \color{gray} 5554 $\pm$ 1046  \\
%\multicolumn{1}{|l|}{$Z^+_{n+1} = Z^+_n + a \times CSFR_n\times(1+z_n)\times Z^-_n$} & 3200 $\pm$ 610 & 4357 $\pm$ 493 & 4375 $\pm$ 475 & 4186 $\pm$ 533 \\
%\multicolumn{1}{|l|}{  \color{gray} $Z^+_{n+1} = Z^+_n + a \times CSFR_n\times(1+z_n)^3\times Z^-_n$} &  \color{gray} 1770 $\pm$ 454 &  \color{gray} 2397 $\pm$ 393 &  \color{gray} 2407 $\pm$ 383 &  \color{gray} 2312 $\pm$ 456\\
\multicolumn{1}{l}{$Z^+_{n+1} = Z^+_n + a \times CSFR_n$} & purple & 5786 $\pm$ 2316 & 7752 $\pm$ 2478 & 7776 $\pm$ 2449 & 7462 $\pm$ 2522   \\
\multicolumn{1}{l}{$Z^+_{n+1} = Z^+_n + a \times CSFR_n\times(1+z_n)$} & blue & 4771 $\pm$ 1535 & 6432 $\pm$ 1571 & 6454 $\pm$ 1547 & 6184 $\pm$ 1601 \\
\multicolumn{1}{l}{$Z^+_{n+1} = Z^+_n + a \times CSFR_n\times(1+z_n)^3$} & cyan & 2818 $\pm$ 532 & \bf  3836 $\pm$ 412 & \bf  3851 $\pm$ 395 & \bf  3687 $\pm$ 466    \\
\multicolumn{1}{l}{$Z^+_{n+1} = Z^+_n + a \times CSFR_n\times Z^-_n$} & yellow & 4266 $\pm$ 1070 & 5785 $\pm$ 1036 & 5807 $\pm$ 1014 & 5554 $\pm$ 1046  \\
\multicolumn{1}{l}{$Z^+_{n+1} = Z^+_n + a \times CSFR_n\times(1+z_n)\times Z^-_n$} & orange & 3200 $\pm$ 610 & \bf  4357 $\pm$ 493 & \bf  4375 $\pm$ 475 & \bf  4186 $\pm$ 533 \\
\multicolumn{1}{l}{ $Z^+_{n+1} = Z^+_n + a \times CSFR_n\times(1+z_n)^3\times Z^-_n$} & red & 1770 $\pm$ 454 & 2397 $\pm$ 393 & 2407 $\pm$ 383 & 2312 $\pm$ 456\\

\hline
\end{tabular}
\end{center}
\vspace{-0.2 cm}
%\caption{\label{ceb_values}
{ $f_{ceb}$ values for the different metallicity evolution cases and the different LGRB rate (function of redshift) estimates of \cite{Lien} and the three models (RF, AB, and NN) of \cite{Graff}.  As expected the cases with poor fitting in Figure \ref{Fitted_rate_plot} / Table \ref{Fitted_rate_values} is also have low accuracy here.  The two best fitted cases are in bold.} 
\end{table*}

In Table \ref{ceb_values} we solve for the $f_{ceb}^{-1}$ value for each of our different metallicity evolution cases, using the LGRB predicted rates from the model of \cite{Lien} and the three machine learning models of \cite{Graff}.  As the shape of the low metallicity star-formation does not perfectly match the LGRB rate estimates, the resulting $f_{ceb}^{-1}$ estimate varies slightly with redshift.  We thus present the average and standard deviation of $f_{ceb}^{-1}$ across the 1 $<$ z $<$ 4 range.  We note that two of the metallicity evolution cases (colored cyan and orange) give errors $\sin$12\%  whereas the other models give errors on the order of 30-40\%.  Using both of these cases across all three \cite{Graff} methods (the bold values in Table \ref{ceb_values}) we estimate a $f_{ceb}^{-1}$ value of 4048 $\pm$ 550.  This is notably higher than the 3135 $\pm$ 668 estimated from the \cite{Lien} numbers for these same two metallicity evolution cases.  (These errors are derived from the values given in Table \ref{ceb_values} only).  Given this discrepancy as well as the 25\% error in estimating the number of Type Ic-bl SNe per M$_\sun$ of star-formation, and the accuracy limitations of the cosmic star-formation rate history itself, we adopt a value of 4000 $\pm$ 2000 as a reasonable estimate %with error 
of our combined efficiency and beaming fraction.

%of 4000 $\pm$ 2000.  Assuming a semi-nominal beaming factor of 200 this suggests that approximately 1 out of 20 low metallically broad-line Type Ic SNe events give rise to an LGRB.

%he slope of the third piece of the CSFR is not well known and at z $<$ 1 the LGRB rates don't have the data size to be sensitive to the different slope of the CSFR).  

\section{Future Work}

There is considerable room for improvement in almost every aspect of this analysis.  Our goal here has been to outline the general methodology and do an initial quick employment of it using information conveniently available in the existing literature.  By far the largest uncertainty is in the beaming factor estimations for LGRBs.  We have attempted to largely sidestep this here by treating the efficiency of forming LGRBs and the degree to which they are beamed as a combined unknown to solve for and allowing he reader to easily use their own preferred beaming estimates.  Obviously a greater understanding of beaming would thus improve our understanding of LGRB production efficiency.  Ultimately an expansion of the \cite{Lien} / \cite{Graff} simulations to all LGRB orientations with proper modeling of the off axis gamma emissions is desirable.  Presumably such an expansion would require an increase of computing effort roughly equivalent to the beaming factor (approximately two orders of magnitude) however with reasonable thresholding prior to detector convolution (i.e. quickly discarding the majority of events obviously too weak for any chance of detection) it might be possible to limit the computational increase to less than an order of magnitude.

A more eminently practical expansion of the \cite{Lien} / \cite{Graff} simulations would be to switch from modeling parameters for arbitrary broken power-law LGRB rates to modeling functional forms more related to the CSFR and metallicity evolution of the universe.  Our current approach of comparing the LGRB event rates estimated by these broken power-law functional forms to our estimated low metallicity CSFR provides a reasonable approximation.  However directly modeling an LGRB event rate from the low metallicity CSFR will likely yield some improvement, as it will be physically motivated.  Such an approach should allow for a better overall fit between the predicted and measured LGRB redshift distribution.

Additionally the existing observational studies of the LGRB redshift distribution from, \cite{wp}, \cite{butler} and the \cite{Fynbo2009} (used in the \citealt{Lien} / \citealt{Graff} simulations), date from 2010 or earlier.  The number of {\it Swift} BAT LGRB detections has since more than doubled.  There are now more then 400 LGRBs with known redshifts.\footnote{\protect\url{http://www.mpe.mpg.de/~jcg/grbgen.html}}  Assuming selection effects could be properly addressed, an analysis of this larger sample would considerably improve the precision on the redshift distribution of bursts which {\it SWIFT} detected and should thus allow a more accurate understanding of the true LGRB redshift distribution both in its shape and scaling.

Further observations of broad-line Type Ic SNe events with and without associated LGRB events could prove particularly insightful especially if distinguishing properties can be identified in the SNe characteristics between these populations.  However any such distinguishing property mush be tested against a low metallicity non-LGRB broad-line Type Ic SNe population specifically.  For example \cite{Cano2013} claim that Ic-bl events with associated LGRBs are more energetic and eject more mass then those Ic-bl SNe without an associated LGRB.  Yet it is unclear if this is due to the non-LGRB events being at typically higher metallicities or due to an intrinsic difference in central engine activity for LGRBs.  The absence of any distinguishable difference in SNe properties with LGRB energy suggests a metallicity effect is more likely.

Finally a better understanding of the distribution of metallicity and its evolution with redshift is essential for furthering this work.  Here we have resorted to testing six very simplistic first principle models for the evolution of the low metallicity star-formation fraction.  Ideally this could be replaced with observational constraints, however a suitable population of uniformly selected high redshift galaxies with measured metallicities is not available and it appears that the the best hope for near-term improvement is with galaxy simulations along the lines of the  L-GALAXIES semi-analytic model \citep{Yates2012}.  Additionally being able to cross-calibrate and thus employ both emission and absorption metallicity diagnostics would greatly assist with tracking the expected eventual convergence of the LGRB and general galaxy population metallicity distribution once redshifts are high enough that the metallicity of the general galaxy population is as low as seen in typical LGRB hosts.  Amusingly LGRBs are themselves likely to be the best events for backlighting absorption features in galaxies with strong emission features. Once the metallicity distributions of the LGRB and general galaxy populations have converged, it should be possible to use LGRBs as a probe of high redshift star-formation (c.f. \citealt{Yuksel}).

\section{Conclusions}

Here we have used the CSFR history \citep{hb}, the core-collapse SNe production rate \citep{SNR007}, and the relative rates of different SNe types \citep{Kelly}, along with estimates of metallicity evolution, distribution, and impact on LGRB production \citep{stats_paper,diff_rate_letter} to derive a equation which estimates the LGRB rate in terms of a remaining unknown constant factor.  This constant factor is a combination of the poorly known LGRB beaming factor and the unknown Ic-bl SN-to-LGRB efficiency factor in ideal low metallicity environments.  We then apply the LGRB rate estimates from simulations of the {\it Swift} BAT detector \citep{Lien, Graff}, which reproduce the observed {\it Swift} LGRB redshift rate and distribution \citep{Fynbo2009,wp}, to estimate that there are 4000 $\pm$ 2000 Type Ic-bl SNe in low-metallicity environments for every LGRB aligned in our direction.  This number is a composite of the fraction of such SNe which produce LGRBs and the fraction that are beamed in our direction.  We have ignored second order complexities such as LGRBs possibly not originating from SNe explosions, the canonical examples typically given as GRBs 060505 and 060614 (however GRB 060614 is generally considered a short burst, \citealt{Dado060614, Bianco060614, Kaneko060614}; etc., and 060505 is also disputed, \citealt{Ofek}), or LGRBs arising from non-broad-line SNe events, as has been claimed for GRB 130215A (SN 2013ez --- \citealt{non_bl_LGRB_SN}).  Thus, assuming a semi-nominal beaming factor of 100, this would suggest that approximately 1 out of 40 low metallically broad-line Type Ic SNe events give rise to an LGRB .

These results are consistent with the absence of off axis LGRB detections in radio surveys of broad-line Type Ic SNe events.  Given the 1 out of 40 rate estimated here a sample size of about a hundred such (low metallicity) SNe would be required to be reasonably confident of detecting an off axis LGRB.  The effect of metallicity on such searches is significant --- a search performed without regard to metallicity will be about 5 times less effective in finding an LGRB.  A search of the literature suggests that no such low metallicity optimized off axis LGRB radio search has yet been conducted and that neither the singe low metallicity broad-line Type Ic SNe in \cite{stats_paper}, SN 2007ce, or another event possibly at such low metallicity, SN 2006nx, have been subjected to such radio observations.  Furthermore none of the broad-line Type Ic SNe with sub-solar metallicities in \cite{stats_paper} were found via a SNe search targeted at specific galaxies, and the median redshift of the non-targeted population z = 0.0584 exceeds the highest redshift of the targeted population.  This suggests that such off axis LGRB searches may be biased toward SNe in high metallicity galaxies and such an off axis search in high metallicity environments will likely be about 30 to 60 times less effective in finding an LGRB (than a search in low metallicity environments).  This  comparative degradation of search efficiency is independent on the beaming factor (however the ideal 1 out of $\sim$40 factor is dependent on beaming angle).

A radio search focusing exclusively on broad-line Type Ic SNe in environments with KK04 \citep{KobulnickyKewley} metallicities of log(O/H)+12 $<$ 8.4 and containing $\sim$90 targets is thus recommended in the strongest possible terms in order to target off-axis LGRBs (with a $\sim$90\% chance of detecting such an event). The current practicality of such an effort is questionable as finding 100 Type Ic-bl SNe will require typing $\sim$4000 SNe  (based on the SNe type fractions given in section \ref{drake} and Table \ref{LGRB_drake} and well as the statistic of \citealt{Weidong_SNe_rates} that $\sim$25\% of local SNe are Type Ia) all of which will need to be found in environments with only the lowest metallicity 10\% of star-formation and close enough to allow the required radio observations.  We estimate approximately 150 Type Ic-bl SNe per year for the entire universe within z $<$ 0.05 of which $\sim$15 will be in such low metallicity environments.  If recent advances in SNe search capabilities (i.e. Pan-STARRS, PTF, and the upcoming LSST) prove capable of finding about half of these events and are matched by a robust photometric SNe typing capability (to bring the needed spectroscopic observations down to a reasonable number) then it should be possible to find an off-axis LGRB within a decade of diligent searching and to place an S/N $\sim$ 3 estimate on the off-axis LGRB rate in a few decades.

\acknowledgments

We thank Andrew Fruchter, Jochen Greiner, Robert Yates, Hendrik van Eerten, David Alexander Kann, and in particular Any Lien for useful discussions.

\bibliographystyle{apj_links}

\bibliography{\jobname}

\end{document}